\documentstyle[12pt,equation,epsfig]{article}

\newcommand{\be}{\begin{equation}}
\newcommand{\ben}{\begin{subequations}}
\newcommand{\een}{\end{subequations}}
\newcommand{\beq}{\begin{eqalignno}}
\newcommand{\eeq}{\end{eqalignno}}
\newcommand{\ee}{\end{equation}}
\newcommand{\wt}{\widetilde}
\newcommand{\imag}{\Im {\rm m}}
\newcommand{\real}{\Re {\rm e}}
\newcommand{\tanb}{\tan \! \beta}
\newcommand{\cotb}{\cot \! \beta}
\newcommand{\mto}{m^2_{\tilde{t}_1}}
\newcommand{\mtt}{m^2_{\tilde{t}_2}}
\newcommand{\mbo}{m^2_{\tilde{b}_1}}
\newcommand{\mbt}{m^2_{\tilde{b}_2}}
\newcommand{\ghat}{\hat{g}^2}
\newcommand{\htop}{\left| h_t \right|^2}
\newcommand{\hb}{\left| h_b \right|^2}
\def\lsim{\:\raisebox{-0.5ex}{$\stackrel{\textstyle<}{\sim}$}\:}
\def\gsim{\:\raisebox{-0.5ex}{$\stackrel{\textstyle>}{\sim}$}\:}

\begin{document}  
\renewcommand{\thefootnote}{\fnsymbol{footnote}}

\pagestyle{empty}
\begin{flushright}
KIAS P00012 \\
TUM-HEP-368-00
\end{flushright}

\vspace{1cm}

\begin{center}

{\large \bf Loop Corrections to the Neutral Higgs Boson Sector of the
MSSM with Explicit CP Violation}

\vspace{1cm}

S.Y. Choi$^1$\footnote{Address after March 1, 2000: Dept. of Physics,
Chonbuk National Univ., Chonju 561--756, Korea}, Manuel Drees$^2$ and
Jae Sik Lee$^1$

\vspace{0.5cm}

{}$^1${\it School of Physics, KIAS, Seoul 130--012, Korea} 

\vspace*{2mm}

{}$^2${\it Physik Dept., TU M\"unchen, James Franck Str., D--85748
Garching, Germany}
\end{center}

\vspace{2cm}

\begin{abstract}

\noindent 
We compute one--loop corrections to the mass matrix of the neutral
Higgs bosons of the Minimal Supersymmetric Standard Model with
explicit CP violation. We use the effective potential method, allowing
for arbitrary splitting between squark masses. We include terms ${\cal
O}(g^2 h^2)$, where $g$ and $h$ stand for electroweak gauge and Yukawa
couplings, respectively. Leading two--loop corrections are taken into
account by means of appropriately defined running quark masses.

\end{abstract}

\newpage

\setcounter{page}{1}
\pagestyle{plain}

\section*{1) Introduction}

Supersymmetry is the currently best motivated extension of the
Standard Model (SM) of particle physics, since it allows to stabilize
the gauge hierarchy without getting into conflict with electroweak
precision data. Among all possible supersymmetric theories, the
Minimal Supersymmetric Standard Model (MSSM) occupies a special
position. It is not only the simplest, i.e. most economical,
potentially realistic supersymmetric field theory, it also has just
the right particle content to allow for the unification of all gauge
interactions \cite{1}. Within the MSSM, the Higgs sector can be
singled out: among all the as yet undetected new particles in the MSSM
spectrum, the lightest neutral Higgs boson $h_1$ is the only one for which a
strict upper bound on the mass can be given \cite{2}, $m_{h_1} \lsim 130$
GeV. A good understanding of the Higgs sector is therefore of crucial
importance for experimental tests of the MSSM.

It has recently been realized \cite{3} that explicit CP violation in
the mass matrices of third generation squarks can induce sizable CP
violation in the MSSM Higgs sector through loop corrections. Note that
CP violating phases for third generation sfermions can be quite large,
since they contribute to the electric dipole moments of the electron
and neutron only at the two--loop level \cite{4}. Although a one--loop
effect, the induced CP violation in the MSSM Higgs sector can be large
enough to significantly affect Higgs phenomenology at present
\cite{3,5} and future \cite{3,6} colliders. An accurate treatment of
CP violating loop corrections to the MSSM Higgs sector is therefore of
some importance.

The first calculations \cite{3} used diagrammatic methods, and
diagonalized the resulting mass matrix only approximately. More
recently, the effective potential has been used \cite{5,7} to compute
the $3 \times 3$ mass matrix of the neutral Higgs bosons of the
MSSM. However, the results of ref.\cite{5} are not valid for large
mass splitting between squark mass eigenstates, while ref.\cite{7}
does not include contributions from the bottom--sbottom sector, which
can be important for large ratio of vacuum expectation values
$\tanb$. Here we present a calculation based on the full one--loop
effective potential, valid for all values of the relevant soft
breaking parameters. This extends older results \cite{8} where CP was
assumed to be conserved. Moreover, unlike refs.\cite{5,7,8} we include
terms ${\cal O}(g^2 h^2)$, where $g$ and $h$ are a weak gauge and
third generation Yukawa coupling, respectively. These new terms do not
change the spectrum very much, but alter CP violating mixing angles by
typically 20\%. Finally, we absorb leading two--loop corrections into
appropriately defined running quark masses \cite{9,2}.

The remainder of this article is organized as follows. In Sec.~2 we
present analytical results for the $3 \times 3$ mass matrix of the
neutral Higgs bosons of the MSSM. In the appropriate limit we find
complete agreement with ref.\cite{5}, but there are some discrepancies
between our results and those of ref.\cite{7}. In Sec.~3 we show some
numerical results, and compare them with results obtained using the
formalism of ref.\cite{5}. Finally, Sec.~4 is devoted to a brief
summary and conclusions.

\section*{2) Analytical results}

The MSSM contains two Higgs doublets $H_1, \ H_2$, with hypercharges
$Y(H_1) = -Y(H_2) = -1/2$. Here we are only interested in the neutral
components, which we write as
\be \label{e1}
H_1^0 = \frac{1} {\sqrt{2}} \left( \phi_1 + i a_1 \right); \ \ \ \ \
H_2^0 = \frac {{\rm e}^{i \xi}} {\sqrt{2}} \left( \phi_2 + i a_2 \right),
\ee
where $\phi_{1,2}$ and $a_{1,2}$ are real fields. The constant phase
$\xi$ can be set to zero at tree level, but will in general become
non--zero once loop corrections are included. 

The mass matrix of the neutral Higgs bosons can be computed from the
effective potential \cite{10}
\beq \label{e2}
V_{\rm Higgs} &= \frac{1}{2} m_1^2 \left( \phi_1^2 + a_1^2 \right)
+ \frac{1}{2} m_2^2 \left( \phi_2^2 + a_2^2 \right) - \left| m_{12}^2
\right| \left( \phi_1 \phi_2 - a_1 a_2 \right) \cos (\xi + \theta_{12})
  \nonumber \\ 
&- \left| m^2_{12} \right| 
\left( \phi_1 a_2 + \phi_2 a_1 \right) \sin ( \xi + \theta_{12})
+ \frac {\hat{g}^2} {8} {\cal D}^2 
+ \frac {1} {64 \pi^2} {\rm Str} \left[
{\cal M}^4 \left( \log \frac { {\cal M}^2} {Q_0^2} - \frac{3}{2}
\right) \right],
\eeq
where we have allowed the soft breaking parameter $m^2_{12} = \left|
m^2_{12} \right| {\rm e}^{i \theta_{12}}$ to be complex. We have
introduced the quantities
\be \label{e5}
{\cal D} = \phi_2^2 + a_2^2 - \phi_1^2 - a_1^2; \ \ \ \ \ \hat{g}^2 =
\frac {g^2 + g'^2} {4} ,
\ee
where the symbols $g$ and $g'$ stand for the $SU(2)$ and $U(1)_Y$
gauge couplings, respectively. $Q_0$ in eq.(\ref{e2}) is the
renormalization scale; the parameters of the tree--level potential, in
particular the mass parameters $m_1^2, \ m_2^2 $ and $m_{12}^2$, are
running parameters, taken at scale $Q_0$. The potential (\ref{e2}) is
then independent of $Q_0$, up to two--loop corrections.

${\cal M}$ is the field--dependent mass matrix of all modes that
couple to the Higgs bosons. The by far dominant contributions come
from third generation quarks and squarks. The (real) masses of the
former are given by
\be \label{e3}
m_b^2 = \frac{1}{2} \hb \left( \phi_1^2 + a_1^2
\right); \ \ \ \
m_t^2 = \frac{1}{2} \htop \left( \phi_2^2 + a_2^2
\right), 
\ee
where $h_b$ and $h_t$ are the bottom and top Yukawa couplings. The
corresponding squark mass matrices can be written as
\ben \label{e4} \beq
{\cal M}_{\tilde t}^2 &= \mbox{$ \left( \begin{array}{cc} 
m^2_{\wt Q} + m_t^2 - \frac{1}{8} \left( g^2 - \frac{g'^2}{3} \right) {\cal D}
&
- h_t^* \left[ A_t^* \left(H_2^0 \right)^* + \mu H_1^0 \right] \\
- h_t \left[ A_t H^0_2 + \mu^* \left( H_1^0 \right)^* \right] &
m^2_{\wt U} + m_t^2 - \frac{g'^2}{6} {\cal D}
\end{array} \right); $} \label{e4a} \\
{\cal M}_{\tilde b}^2 &= \mbox{$ \left( \begin{array}{cc} 
m^2_{\wt Q} + m_b^2 + \frac{1}{8} \left( g^2 + \frac{g'^2}{3} \right) {\cal D}
&
- h_b^* \left[ A_b^*  \left( H_1^0 \right)^* + \mu H_2^0 \right] \\
- h_b \left[ A_b H_1^0 + \mu^* \left( H_2^0 \right)^* \right] &
m^2_{\wt D} + m_b^2 + \frac{g'^2}{12} {\cal D}
\end{array} \right). $} \label{e4b}
\eeq \een
Here, $H_1^0$ and $H_2^0$ are given by eqs.(\ref{e1}) while $m_t^2$
and $m_b^2$ are as in eqs.(\ref{e3}) and ${\cal D}$ has been defined
in eqs.(\ref{e5}). In eqs.(\ref{e4}) $m^2_{\wt Q}, \ m^2_{\wt U}$ and
$m^2_{\wt D}$ are real soft breaking parameters, $A_b$ and $A_t$ are
complex soft breaking parameters, and $\mu$ is the complex
supersymmetric Higgs(ino) mass parameter. The eigenvalues of the mass
matrices (\ref{e4}) are:
\ben \label{e6} \beq
\label{e6a}
m^2_{\tilde{t}_{1,2}} &= \frac{1}{2} \left[ m^2_{\wt Q} + m^2_{\wt U}
+ \htop \left( \phi_2^2 + a_2^2 \right) -
\frac {\ghat} {2} {\cal D} 
\right. \\ \nonumber & \left. \hspace*{.8cm}
\mp \left\{ \left( m^2_{\wt Q} - m^2_{\wt U} - \frac {3 g^2 - 5 g'^2}
{24} {\cal D} \right)^2 + 2 \htop \left| A_t {\rm e}^{i
\xi} \left( \phi_2 + i a_2 \right) + \mu^* \left( \phi_1 - i a_1
\right) \right|^2 \right\}^{1/2} \right];
 \\
m^2_{\tilde{b}_{1,2}} &= \frac{1}{2} \left[ m^2_{\wt Q} + m^2_{\wt D}
+ \hb \left( \phi_1^2 + a_1^2 \right) +
\frac{\ghat}{2} {\cal D} 
\right. \\ \nonumber & \left. \hspace*{.8cm}
\mp \left\{ \left( m^2_{\wt Q} - m^2_{\wt D} + \frac {3 g^2 - g'^2}
{24} {\cal D} \right)^2 + 2 \hb \left| A_b 
\left( \phi_1 + i a_1 \right) + \mu^* {\rm e}^{-i\xi} \left( \phi_2 - i a_2
\right) \right|^2 \right\}^{1/2} \right].
\label{e6b}
\eeq \een

The calculation proceeds by plugging the field--dependent eigenvalues
(\ref{e3}) and (\ref{e6}) into the potential (\ref{e2}); here the
complex scalar squarks and Dirac fermion quarks enter with overall
factors $+2$ and $-4$, respectively. The mass matrix of the Higgs
bosons (at vanishing external momentum) is given by the matrix of
second derivatives of this potential, computed at its minimum. In
order to make sure that we are indeed in the minimum of the potential,
we solve the stationarity relations, i.e. set the first derivatives of
the potential to zero. This allows us to, e.g., express $m_1^2, \
m_2^2$ and $m_{12}^2 \sin (\xi +\theta_{12})$ as functions of the
vacuum expectation values (vevs) and the remaining parameters
appearing in the loop--corrected Higgs potential. Note that the
equations $\partial V_{\rm Higgs} / \partial a_1 = 0$ and $\partial
V_{\rm Higgs} / \partial a_2 = 0$ are linearly dependent, i.e. lead to
only one constraint on parameters, if we demand that $\langle a_1
\rangle = \langle a_2 \rangle = 0$; the remaining vevs are defined
through
\be \label{e6nn}
\langle \phi_1 \rangle^2 + \langle \phi_2 \rangle^2 = \frac {M_Z^2}
{\ghat} \simeq (246 \ {\rm GeV})^2; \ \ \frac{ \langle \phi_2 \rangle}
{\langle \phi_1 \rangle} = \tanb.
\ee

This leads to the following expression for the re--phasing invariant
sum $\xi + \theta_{12}$:
\be \label{e7}
m_{12}^2 \sin (\xi + \theta_{12}) 
= -\frac {3} {32 \pi^2} \left\{ \left[ f(\mto) -
f(\mtt) \right] \htop \Delta_{\tilde t} + \left[
f(\mbo) - f(\mbt) \right] \hb \Delta_{\tilde b} \right\},
\ee
where
\be \label{e8}
f(m^2) = 2 m^2 \left( \log \frac {m^2} {Q_0^2} - 1 \right).
\ee
In eq.(\ref{e7}) we have introduced the quantities
\be \label{e9}
\Delta_{\tilde t} = \frac { \imag(A_t \mu {\rm e}^{i \xi}) }
{\mtt - \mto} ; \ \
\Delta_{\tilde b} = \frac { \imag(A_b \mu {\rm e}^{i \xi}) }
{\mbt - \mbo} ,
\ee
which describe the amount of CP violation in the squark mass
matrices. Note that $\Delta_{\tilde t}$ remains finite as $\mto
\rightarrow \mtt$, since this implies $\imag(A_t \mu {\rm e}^{i \xi})
\rightarrow 0$. Most tree--level analyses use the convention $\xi=0$;
in this case we could set it to zero in the right--hand side of
eq.(\ref{e7}), to one--loop order. We kept it in order to illustrate
that only the phases of the re--phasing invariant quantities $A_t \mu
{\rm e}^{i \xi}$ and $A_b \mu {\rm e}^{i \xi}$ have physical meaning.

The mass matrix of the neutral Higgs bosons can now be computed from
the matrix of second derivatives of the potential (\ref{e2}), where
(after taking the derivatives) $m_1^2, \ m_2^2$ and $m_{12}^2 \sin(
\xi + \theta_{12})$ are determined by the stationarity conditions. We
find that the state $G^0 = a_1 \cos \beta - a_2 \sin \beta$ is
massless; it describes the would--be Goldstone mode that gets
``eaten'' by the longitudinal $Z$ boson. We are thus left with a
squared mass matrix ${\cal M}_H^2$ for the three states $a = a_1 \sin
\beta + a_2 \cos \beta, \ \phi_1$ and $\phi_2$. This matrix is real
and symmetric, i.e. it has 6 independent entries. The diagonal entry
for $a$ reads:
\be \label{e10}
\left. {\cal M}^2_{H} \right|_{aa} = m_A^2 + \frac {3} {8 \pi^2}
\left\{ \frac { \htop m_t^2 } { \sin^2 \beta} g(\mto,
\mtt) \Delta_{\tilde t}^2 + 
\frac { \hb m_b^2 } { \cos^2 \beta} g(\mbo,
\mbt) \Delta_{\tilde b}^2 \right\},
\ee
where $\Delta_{\tilde t}$ and $\Delta_{\tilde b}$ are as in
eq.(\ref{e9}), and we have introduced the function
\be \label{e11}
g(m_1^2, m_2^2) = 2 - \frac {m_1^2 + m_2^2} {m_1^2 - m_2^2} \log \frac
{m_1^2} {m_2^2} .
\ee
The quantity $m_A^2$ in eq.(\ref{e10}) is given by
\be \label{e12}
m_A^2 = \frac {2 m_{12}^2 \cos (\xi + \theta_{12})} {\sin(2\beta)} 
+ \frac {2} {\sin(2 \beta)} \left\{ \htop \real (A_t \mu
{\rm e}^{i \xi} ) F(\mto,\mtt) + \hb \real (A_b \mu {\rm e}^{i \xi} ) 
F(\mbo,\mbt) \right\},
\ee
where
\be \label{en1}
F(m_a^2,m_b^2) = \frac {3} {32 \pi^2} \frac {f(m_a^2)-f(m_b^2)} 
{m_b^2-m_a^2};
\ee
the function $f$ has been defined in eq.(\ref{e8}).

If we consider $m_A^2$, the values of the soft breaking parameters
and $\mu$ to be inputs, eqs.(\ref{e7}) and (\ref{e12}) can be combined
to give an explicit expression for the induced phase $\xi$ of the vev:
\be \label{en2}
\sin \xi = -\frac { \htop \imag (A_t \mu) F(\mto,\mtt) + \hb \imag (A_b
\mu) F(\mbo,\mbt) + \imag( m^2_{12}) }
{ m_A^2 \sin \beta \cos \beta }.
\ee
It should be emphasized that $\xi$ and $\theta_{12}$ are not
separately physical quantities; only their sum is re--phasing
invariant. Nevertheless eq.(\ref{en2}) can be useful. Note that the
leading $Q_0$ dependence in the numerator cancels between the explicit
terms $\propto \log Q_0$ contained in $F(\mto,\mtt)$ and
$F(\mbo,\mbt)$, and the RG--induced running of $\imag (m^2_{12})$. If
one sets the phase of the Higgs field at some scale, eq.(\ref{en2})
shows that it will remain scale--independent, at least to 1--loop
order. In particular, $\xi$ can be set to zero, as pointed out by
Pilaftsis \cite{3}. Eq.(\ref{en2}) then determines the loop--induced
phase of $m_{12}^2$ (called a counter--term in ref.\cite{3}). On the
other hand, it is also possible to set $\theta_{12} = 0$. The price
one has to pay is that the phase of the Higgs fields will show a
strong scale dependence even after explicit 1--loop corrections to the
Higgs potential have been added.

We will see below that most of the $Q_0$ dependence of the loop
corrections to the Higgs mass matrix can be absorbed into the
parameter $m_A$. Within $m_A^2$ itself, the $Q_0$ dependence largely
cancels between the running of $m_{12}^2$ and the explicit $\log Q_0$
dependence of $f(m^2)$, eq.(\ref{e8}), just as in the CP conserving
case \cite{8}. Note that our $m_A$ differs from $M_a$ defined in
ref.\cite{5}, since our $m_A$ only includes corrections that are
nonzero in the limit of exact CP invariance, whereas $M_a^2$
corresponds to our $\left. {\cal M}_H^2\right|_{aa}$ of
eq.(\ref{e10}). In the remainder of this paper we will consider
$m_A^2$ as well as the re--phasing invariant quantities $A_t \mu {\rm
e}^{i\xi}$ and $A_b \mu {\rm e}^{i\xi}$ to be input parameters. The
results are then independent of the convention adopted for
$\theta_{12}$. 

The CP violating entries of the mass matrix, which mix $a$ with
$\phi_1$ and $\phi_2$, are:
\ben \label{e13} \beq
\label{e13a} 
\left. {\cal M}^2_H \right|_{a \phi_1} &= \frac {3} {16 \pi^2} \left\{
\frac { m_t^2 \Delta_{\tilde t} } {\sin \beta} \left[ g(\mto, \mtt)
\left( X_t \cotb - 2 \htop R_t \right) - \ghat \cotb \log \frac
{\mtt} {\mto} \right]
\right. \\ & \left. \hspace*{15mm}
+ \frac {m_b^2 \Delta_{\tilde b}} {\cos \beta} \left[ -g(\mbo,\mbt)
\left( X_b + 2 \hb R_b' \right) + \left( \ghat - 2 \hb \right) \log
\frac {\mbt} {\mbo} \right] \right\};
\nonumber \\
\label{e13b}
\left. {\cal M}^2_H \right|_{a \phi_2} &= \frac {3} {16 \pi^2} \left\{
\frac {m_t^2 \Delta_{\tilde t}} {\sin \beta} \left[ -g(\mto,\mtt)
\left( X_t + 2 \htop R_t' \right) + \left( \ghat - 2 \htop \right) \log
\frac {\mtt} {\mto} \right]
\right. \\ & \left. \hspace*{15mm}
+ \frac { m_b^2 \Delta_{\tilde b} } {\cos \beta} \left[ g(\mbo, \mbt)
\left( X_b \tanb - 2 \hb R_b \right) - \ghat \tanb \log \frac
{\mbt} {\mbo} \right]  \right\}. \nonumber
\eeq \een
As noted earlier, the size of these entries is controlled by
$\Delta_{\tilde t}$ and $\Delta_{\tilde b}$. In addition we have
introduced the dimensionless quantities
\ben \label{e14} \beq
X_t & = \frac{ 5 g'^2 - 3 g^2} {12} \cdot \frac {m^2_{\wt Q} - m^2_{\wt U}}
{\mtt - \mto}; \ \ \ X_b = \frac{ g'^2 - 3 g^2} {12} \cdot 
\frac {m^2_{\wt Q} - m^2_{\wt D}} {\mbt - \mbo};
\label{e14a} \\
R_t & = \frac { \left| \mu \right|^2 \cotb + \real(A_t \mu {\rm e}^{i
\xi})} { \mtt - \mto} ; \ \ \
R_t' = \frac { \left| A_t \right|^2 + \real(A_t \mu {\rm e}^{i\xi})
\cotb} { \mtt - \mto} ;
\label{e14b} \\
R_b & = \frac { \left| \mu \right|^2 \tanb + \real(A_b \mu {\rm e}^{i
\xi})} { \mbt - \mbo} ; \ \ \
R_b' = \frac { \left| A_b \right|^2 + \real(A_b \mu {\rm e}^{i\xi})
\tanb} { \mbt - \mbo} .
\label{e14c} 
\eeq \een
As in eqs.(\ref{e9}), phases only appear in the re--phasing invariant
combinations $A_t \mu {\rm e}^{i \xi}$ and $A_b \mu {\rm e}^{i
\xi}$. The terms proportional to $\ghat$ [defined in eqs.({\ref{e5})],
$X_t$ or $X_b$ are mixed gauge--Yukawa contributions, which were
neglected in refs.\cite{5,7}; note that no corrections of this kind
appear in eqs.(\ref{e7}) and (\ref{e10}). We do not include pure
gauge, ${\cal O}(g^4)$ corrections, since there are many additional
corrections of this order from first and second generation sfermions
as well as from loops involving gauge and Higgs bosons and their
superpartners. 

While eqs.(\ref{e13}) are pure loop corrections, the remaining entries
of the Higgs boson mass matrix also receive tree--level contributions:
\ben \label{e15} \beq
\left. {\cal M}^2_H \right|_{\phi_1 \phi_1} &=
M_Z^2 \cos^2 \beta + m_A^2 \sin^2 \beta
\nonumber \\ &
+ \frac {3 m_t^2} {8 \pi^2} \left[ g(\mto,\mtt) R_t \left( \htop R_t -
\cotb X_t \right) + \ghat \cotb R_t \log \frac {\mtt} {\mto} \right]
\nonumber \\ &
+ \frac {3 m_b^2} {8 \pi^2} \left\{ \hb \log \frac {\mbo \mbt} {m_b^4}
- \ghat \log \frac {\mbo \mbt} {Q_0^4}
\right. \label{e15a} \\ & \left. \hspace*{14mm}
+ g(\mbo,\mbt) R_b' \left( \hb R_b' + X_b \right) + \log \frac {\mbt}
{\mbo} \left[ X_b + \left( 2 \hb - \ghat \right) R_b' \right]
\right\};
\nonumber \\
\left. {\cal M}_H^2 \right|_{\phi_1 \phi_2} & = 
- \left( M_Z^2 + m_A^2 \right) \sin \beta \cos \beta
\nonumber \\ &
+ \frac {3 m_t^2} {8 \pi^2} \left\{ g(\mto,\mtt) \left[ \htop R_t R_t'
+ \frac {X_t}{2} \left( R_t - R_t' \cotb \right) \right] + \frac
{\ghat} {2} \cotb \log \frac { \mto \mtt} {Q_0^4}
\nonumber \right. \\ & \left. \hspace*{14mm}
+ \log \frac {\mtt} {\mto} \left[ \htop R_t - \frac {X_t}{2} \cotb +
\frac {\ghat}{2} \left( R_t' \cotb - R_t \right) \right] \right\}
\nonumber \\ &
+ \frac {3 m_b^2} {8 \pi^2} \left\{ g(\mbo,\mbt) \left[ \hb R_b R_b'
+ \frac {X_b}{2} \left( R_b - R_b' \tanb \right) \right] + \frac
{\ghat} {2} \tanb \log \frac { \mbo \mbt} {Q_0^4}
\nonumber \right. \\ & \left. \hspace*{14mm}
+ \log \frac {\mbt} {\mbo} \left[ \hb R_b - \frac {X_b}{2} \tanb +
\frac {\ghat}{2} \left( R_b' \tanb - R_b \right) \right] \right\};
\label{e15b} \\
\left. {\cal M}^2_H \right|_{\phi_2 \phi_2} &=
M_Z^2 \sin^2 \beta + m_A^2 \cos^2 \beta
\nonumber \\ &
+ \frac {3 m_t^2} {8 \pi^2} \left\{ \htop \log \frac {\mto \mtt} {m_t^4}
- \ghat \log \frac {\mto \mtt} {Q_0^4}
\right. \nonumber \\ & \left. \hspace*{14mm}
+ g(\mto,\mtt) R_t' \left( \htop R_t' + X_t \right) + \log \frac {\mtt}
{\mto} \left[ X_t + \left( 2 \htop - \ghat \right) R_t' \right]
\right\}
\nonumber \\ &
+ \frac {3 m_b^2} {8 \pi^2} \left[ g(\mbo,\mbt) R_b \left( \hb R_b -
\tanb X_b \right) + \ghat \tanb R_b \log \frac {\mbt} {\mbo} \right] .
\label{e15c}
\eeq \een
As noted earlier, the explicit dependence on the renormalization scale
$Q_0$ has mostly been absorbed into $m_A^2$. The only exceptions are
terms proportional to the combination $\ghat$ of electroweak gauge
couplings, eqs.(\ref{e5}). They come from the wave function
renormalization of the Higgs fields, which leads to a logarithmic
$Q_0$ dependence of the tree--level contributions $\ghat \langle
\phi_1 \rangle^2$ (in $\left. {\cal M}^2_H \right|_{\phi_1 \phi_1}$),
$- \ghat \langle \phi_1 \rangle \langle \phi_2 \rangle$ (in 
$\left. {\cal M}^2_H \right|_{\phi_1 \phi_2}$) and $\ghat \langle
\phi_2 \rangle^2$ (in $\left. {\cal M}^2_H \right|_{\phi_2 \phi_2}$),
respectively. Following ref.\cite{5} we define these vevs at scale
$Q_0 = m_t$, i.e. we use $Q_0 = m_t$ in the explicitly $Q_0$ dependent
terms in eqs.(\ref{e15}).

In order to compare our results with those of Demir \cite{7}, we have
to set $g=g'=h_b=0$ in the loop corrections. Our results then become
similar to his if we identify our $m_A^2$ with his
$\wt{M}_A^2$. However, his CP--odd quantity $\Delta$ seems to be too
small by a factor of 2, and the coefficient of $g(\mto,\mtt)$ in
$\left. {\cal M}^2_H \right|_{\phi_1 \phi_2}$ differs if CP is
violated.

In order to compare our results with those of Pilaftsis and Wagner
\cite{5}, we have to set most corrections involving gauge interactions
to zero, the only exception being the terms $\propto \ghat \log
Q_0$. In addition, we have to take the limit of small squark mass
splitting, which implies
\ben \label{e16} \beq
\frac {f (m^2_{\tilde{q}_1}) - f(m^2_{\tilde{q}_2}) }
{m^2_{\tilde{q}_1} - m^2_{\tilde{q}_2} } & \longrightarrow 2 \log \frac
{M^2_{\rm SUSY}} {Q_0^2} ;
\label{e16a} \\
\frac { g(m^2_{\tilde{q}_1}, m^2_{\tilde{q}_2} ) }
{ \left( m^2_{\tilde{q}_1} - m^2_{\tilde{q}_2} \right)^2 } & \longrightarrow
- \frac {1} {6 M^4_{\rm SUSY}} ;
\label{e16b} \\
\frac {1} {m^2_{\tilde{q}_1} - m^2_{\tilde{q}_2} } \log \frac
{m^2_{\tilde{q}_1}} {m^2_{\tilde{q}_2}} & \longrightarrow \frac {1}
{M^2_{\rm SUSY}},
\label{e16c}
\eeq \een
where $M^2_{\rm SUSY} = \left( m^2_{\tilde{q}_1} + m^2_{\tilde{q}_2}
\right)/2$. After these replacements we find complete agreement with
the results of ref.\cite{5}, except for $\left. {\cal M}^2_H
\right|_{aa}$; however, this difference can be absorbed into the input
parameter $m_A^2$, i.e. it has no physical significance.

Pilaftsis and Wagner also include logarithmically enhanced two--loop
corrections ${\cal O}(g_s^2 h^4)$ and ${\cal O}(h^6)$, where $g_s$ is
the strong gauge coupling. As shown in refs.\cite{9,2}, leading QCD
corrections can be included by interpreting the quark masses and
Yukawa couplings appearing in eqs.(\ref{e10}), (\ref{e12}),
(\ref{e13}) and (\ref{e15}) to be running parameters. For example,
\be \label{e17}
\overline{m}_t(Q) = \overline{m}_t(m_t) \cdot \left[ \frac {g_s^2(Q)}
{g_s^2(m_t)} \right]^{12/21},
\ee
where we have assumed $Q \leq M_{\rm SUSY}$, i.e. only standard QCD
contributes; a pole mass $m_t^{\rm pole} = 173$ GeV corresponds to
$\overline{m}_t(m_t) = 165$ GeV. The corresponding expression for the
running Yukawa coupling follows from eq.(\ref{e3}) and the observation
that, to one--loop order, QCD corrections do not contribute to the
running of $\langle \phi_2 \rangle$. Corrections ${\cal O}(h^6)$ to
the Higgs mass matrix come from the one--loop running of the Yukawa
couplings, as well as (for $Q \leq M_{\rm SUSY}$) from the two--loop
running of the quartic Higgs coupling(s) in the effective
non--supersymmetric theory. Finally, as pointed out in ref.\cite{2},
one can absorb significant ${\cal O}(g_s^2 h^4)$ corrections that are
not enhanced by large logarithms by including gluino--stop loop
corrections to the running top mass at scale $Q \simeq M_{\rm
SUSY}$. In order to (approximately) include all these corrections, we
introduce three different top masses that appear in various
contributions to ${\cal M}^2_H$:
\begin{itemize}
\item
The leading--log term $\propto \htop m_t^2 \log \frac {\mto \mtt}
{m_t^4}$ should be computed with a running top mass at the
intermediate scale $Q^2_{\rm int} =  m_t M_{\rm SUSY}$:
\be \label{e18}
m_{t,{\rm int}} = \overline{m}_t(Q_{\rm int}) \left[ 1 + \frac {3
\htop} {64 \pi^2} \log \frac {Q^2_{\rm int}} {m_t^2} \right].
\ee
\item
In terms proportional to powers of the CP--odd quantity
$\Delta_{\tilde t}$, one should use the top mass
\be \label{e19}
m_{t,{\rm odd}} = m_{t,{\rm high}}  \left[ 1 + \frac {6
\htop} {64 \pi^2} \log \frac {M^2_{\rm SUSY}} {m_t^2} \right].
\ee
\item
Finally, in the remaining terms the top mass should be interpreted as
\be \label{e20}
m_{t,{\rm even}} = m_{t,{\rm high}}  \left[ 1 + \frac {3
\htop} {64 \pi^2} \log \frac {M^2_{\rm SUSY}} {m_t^2} \right].
\ee
\end{itemize}
The quantity $m_{t,{\rm high}}$ appearing in eqs.(\ref{e19},\ref{e20})
is given by
\be \label{e21}
m_{t,{\rm high}} = \overline{m}_t(M_{\rm SUSY}) - \frac {g_s^2}{12
\pi^2} \sin(2 \theta_{\tilde t}) \real( {\rm e}^{-i\phi_{\tilde t}} \, 
m_{\tilde g} )
\left[ B_0(m_t,m_{\tilde{t}_1},|m_{\tilde g}|) - 
 B_0(m_t,m_{\tilde{t}_2},|m_{\tilde g}|) \right].
\ee
Here $\theta_{\tilde t}$ and $\phi_{\tilde t}$ are the angles needed
to diagonalize the stop mass matrix (\ref{e4a}), i.e. $\tilde{t}_1 =
\cos \theta_{\tilde t} \, \tilde{t}_L + \sin \theta_{\tilde t} \, {\rm
e}^{-i \phi_{\tilde t}} \, \tilde{t}_R$. $m_{\tilde g}$ is the gluino
mass, which can be complex, and $B_0$ is the Passarino--Veltman
two--point function. Eq.(\ref{e21}) generalizes corresponding results
of ref.\cite{2} to the case with CP--odd phases; note that the
quantity ${\rm e}^{-i\phi_{\tilde t}} m_{\tilde g}$ is re--phasing
invariant. Eqs.(\ref{e18})--(\ref{e21}) can be extended
straightforwardly to the bottom--sbottom sector.\footnote{In
eq.(\ref{e21}) we have assumed that $m_t < m_{\tilde g} +
m_{\tilde{t}_1}$. If this is not the case, the loop functions will
develop imaginary parts. To one--loop order, only the real part of the
correction contributes.}

\section*{3) Numerical examples}

We are now ready to present some numerical results. It is clear from
eqs.(\ref{e13}) and (\ref{e9}) that loop--induced CP violation in the
Higgs sector can only be large if both $|\mu|$ and $|A_t|$ (or
$|A_b|$, if $\tanb \gg 1$) are sizable \cite{3}. We therefore choose
$|A_t| = |A_b| = |\mu| = 2 m_{\wt Q}$. For definiteness we only
present results for fixed $\tanb=10$, and real and positive gluino
mass. The value of $\tanb$ is not very important, unless it reaches
$\sim m_t/m_b$, where $b - \tilde{b}$ contributions can be
important. Since for the given moderate value of $\tanb$ contributions
from the (s)bottom sector are still quite small, our result are not
sensitive to $m_{\wt D}$ and $A_b$; we therefore fix $m_{\wt D} =
m_{\wt U}$ and also take equal phases for $A_t$ and $A_b$. However, we
allow different values for the soft breaking masses of $SU(2)$ doublet
and singlet squarks, $m_{\wt Q} \ne m_{\wt U}$. This allows us to
study scenarios with large $\tilde{t}_1 - \tilde{t}_2$ mass difference
even if $m_{\tilde Q} \gg m_t$. Note that renormalization group
effects do in fact produce significant differences between $m_{\wt Q}$
and $m_{\wt U}$ even if they are equal at some very large energy scale
\cite{11}. Finally, we take $\overline{m}_t(m_t) = 165$ GeV and
$\overline{m}_b(m_b) = 4.2$ GeV.

We first study a scenario with $m_{\wt Q} = m_{\wt U}$. Here we expect
good agreement with results of ref.\cite{5} as long as $m_{\wt Q} \gg
m_t$, if we ``switch off'' the new contributions $\propto g^2 h^2$
presented in Sec.~2 as well as the SUSY--QCD correction
(\ref{e21}). This expectation is borne out by Fig.~1, which shows the
mass of the lightest neutral Higgs boson (left panel) as well as the
CP violating Higgs mixing angle $\alpha_2$ (right panel) as a function
of $m_{\wt Q}$, for $m_A = 1$ TeV and $\theta_{\rm eff} \equiv
\arg(A_t \mu {\rm e}^{i\xi}) = \pi/2$. Defining the orthogonal Higgs
mixing matrix $O^H$ through $h_i = O^H_{ij} \phi_j$, where $h_i$
denotes the physical Higgs bosons (mass eigenstates) and we have set
$\phi_3 \equiv a$, the $\alpha_i$ are given by
\be \label{e22}
\alpha_i = \min \left[ \frac { \left| O^H_{i3} \right|}
{\sqrt{ \left| O^H_{i1} \right|^2 + \left| O^H_{i2} \right|^2 } }, \ 
\frac {\sqrt{ \left| O^H_{i1} \right|^2 + \left| O^H_{i2} \right|^2 } }
{ \left| O^H_{i3} \right|} \right] .
\ee
If CP is conserved, $h_i$ is either purely CP--odd ($O^H_{i3}=1$), or 
purely CP--even ($\left| O^H_{i1} \right|^2 + \left| O^H_{i2}
\right|^2 = 1$), in which case $\alpha_i = 0$.

\begin{figure}[htb]
\vspace*{-20mm}
\epsfig{file=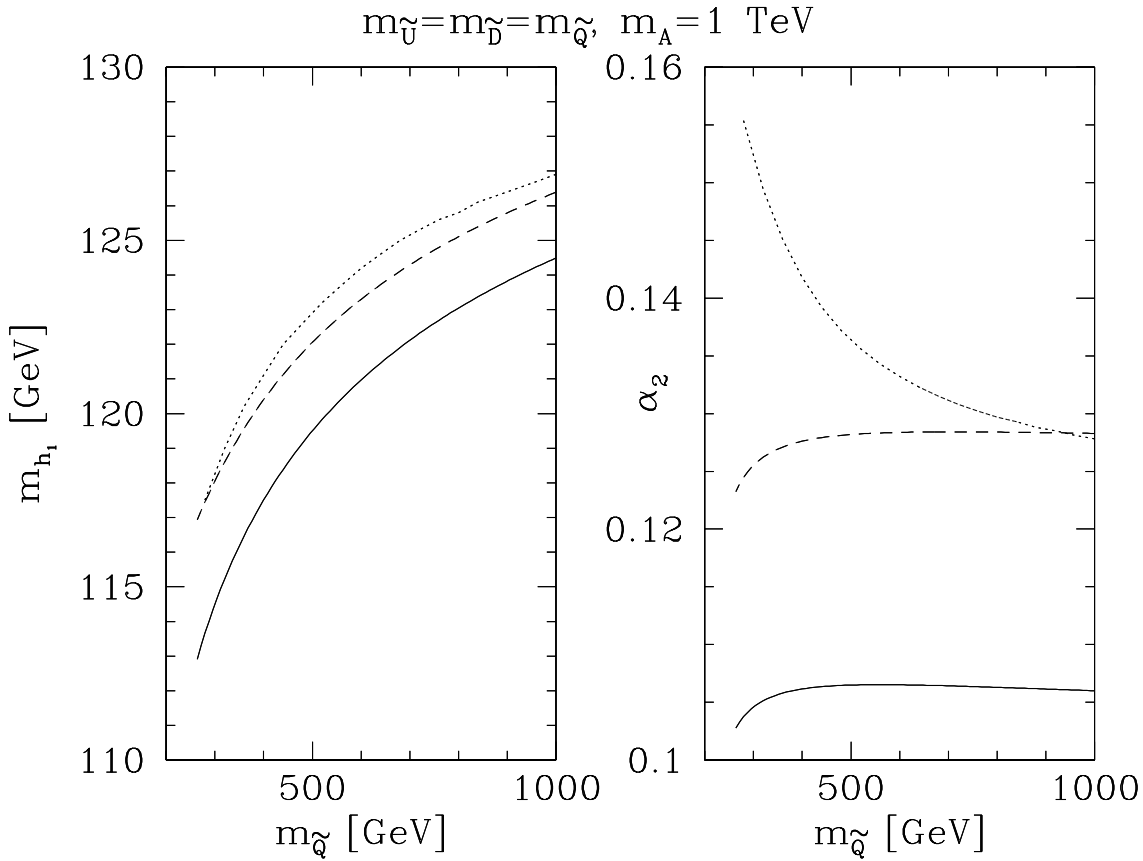,width=0.72\textwidth} 
\vspace*{-14mm}
\caption{\small The mass of the lightest neutral Higgs boson (left) and the
CP violating mixing angle of the second--lightest neutral Higgs boson
(right) as a function of the soft breaking mass $m_{\wt Q}$ of $SU(2)$
doublet squarks. The values of the other free parameters are: $m_{\wt
U} = m_{\wt D} = m_{\wt Q}, \ |A_t| = |A_b| = |\mu| = 2 m_{\wt Q}, \
\tanb=10, \ m_A = 1$ TeV, $m_{\tilde g} = 1$ TeV and squark phase
$\theta_{\rm eff} = \pi/2$. The dotted lines have been obtained using
the expressions of ref.\cite{5}, while the dashed and solid curves
show our results without and with ${\cal O}(g^2 h^2)$ and 2--loop
SUSY--QCD corrections, respectively. }
\end{figure}

The dotted curves in Fig.~1 have been obtained using the expressions
of ref.\cite{5}. As expected, for large $m_{\wt Q}$ we find very good
agreement with the dashed curves, which show our results with the new
${\cal O}(g^2 h^2)$ and 2--loop SUSY--QCD contributions switched
off. For this comparison we have chosen the parameter $M^2_a$ of
ref.\cite{5} to agree with the $(a,a)$ element (\ref{e10}) of the
Higgs mass matrix. We then find that the expressions of ref.\cite{5}
reproduce the mass of the lightest neutral Higgs boson very well even
for quite small values of $m_{\wt Q}$. This is somewhat surprising,
since for the smallest experimentally allowed value of $m_{\wt Q}$,
defined through the requirement $m_{\tilde{t}_1} \geq 90$ GeV
\cite{12}, the stop mass splitting is very large, $m_{\tilde{t}_2}
\simeq 4 m_{\tilde{t}_1}$. On the other hand, the two predictions for
$\alpha_2$ do start to deviate significantly for $m_{\wt Q} \lsim 500$
GeV, which corresponds to $m_{\tilde{t}_2} \gsim 2
m_{\tilde{t}_1}$. Note that $\alpha_1$ is very small in this example,
${\cal O}(10^{-3})$, due to the large difference between $m_A$ and
$m_{h_1}$. 

The solid curves show our full results, including all contributions
described in Sec.~2; we choose a gluino mass of 1 TeV. The new
contributions to $m_{h_1}$ are not very large, although the reduction
by 4 GeV for the smallest allowed value of $m_{\wt Q}$ is not entirely
negligible. In contrast, these new contributions reduce $\alpha_2$ by
$\sim 20\%$ for all values of $m_{\wt Q}$. Contributions of this
magnitude should certainly be taken into account when translating
observed (bounds on) CP violating effects into (bounds on the) values
of the fundamental soft breaking parameters.

\begin{figure}[htb]
\vspace*{-20mm}
\epsfig{file=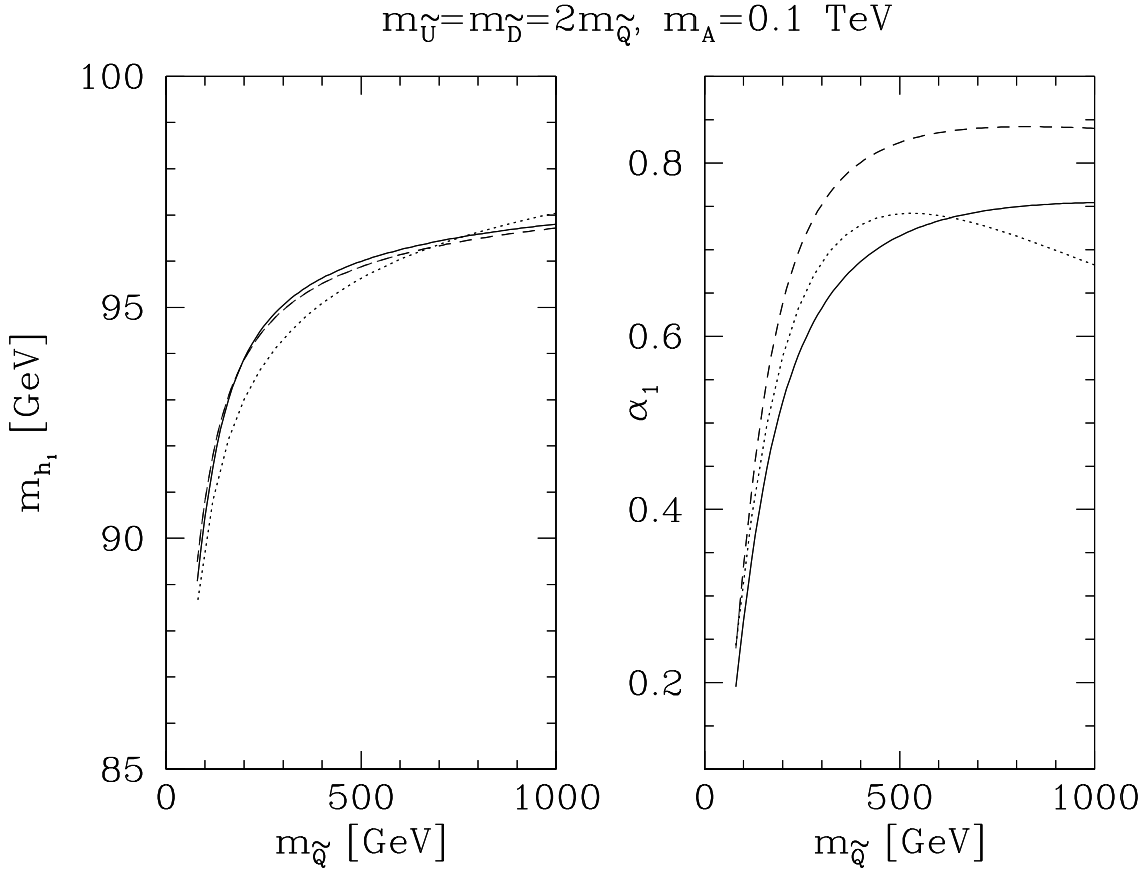,width=0.72\textwidth} 
\vspace*{-14mm}
\caption{\small The mass of the lightest neutral Higgs boson (left) and the
CP violating mixing angle of this Higgs boson (right) as a function of
$m_{\wt Q}$. Notation and parameters are as in Fig.~1, except that we
have taken $m_{\wt U} = m_{\wt D} = 2 m_{\wt Q}$ and $m_A = 100$ GeV. }
\end{figure}

In Fig.~2 we show results for a scenario with $m_{\wt U} = 2 m_{\wt
Q}$, so that $m_{\tilde{t}_2} > 2 m_{\tilde{t}_1}$ for all values of
$m_{\wt Q}$. We nevertheless find good agreement between our
calculation of $m_{h_1}$ and the results of ref.\cite{5}. The reason
is that we now have chosen $m_A = 100$ GeV, in the range of interest
to current LEP experiments. In the absence of CP violation we would
then have $m_{h_1} \simeq m_A$ \cite{8}, since $\tan^2 \beta \gg 1$ in
our example. In the presence of CP violation in the squark sector the
corrections (\ref{e10}) to the $(a,a)$ element of the Higgs mass
matrix are negative; $m_{h_1}$ is reduced further by mixing between
CP--even and CP--odd Higgs bosons. However, these effects remain quite
small as far as the Higgs spectrum is concerned, and therefore need
not be computed with very high accuracy.

In contrast, the three predictions of $\alpha_1$ now differ
significantly over the entire allowed range of $m_{\wt Q}$. In the
given case we observe a partial cancellation between effects due to
squark mass splitting and the new ${\cal O}(g^2 h^2)$ and 2--loop
SUSY--QCD contributions. As a result the prediction from ref.\cite{5}
agrees more closely with our full result than with the approximation
that only includes the interactions also included in ref.\cite{5}. We
see that the new mixed gauge--Yukawa and SUSY--QCD corrections again
reduce CP violation in the Higgs sector by 10 to 20\%. Note, however,
that this loop--induced CP violation in the Higgs sector can still be
very large, i.e. $\alpha_1 \sim {\cal O}(1)$ is possible here.

Finally, in Fig.~3 we show that $\alpha_1$ can in fact reach its
theoretical maximum of unity. In this figure we keep all dimensionful
parameters fixed, and show $m_{h_1}$ and $\alpha_1$ as a function of
the phase $\theta_{\rm eff}$ appearing in the squark mass
matrices. The dependence of $m_{h_1}$ on this phase is mild. Note that
the lightest Higgs mass now takes its maximal value for maximal CP
violation in the squark sector, $\theta_{\rm eff} = \pi/2$. In
contrast, the CP violating Higgs mixing angle $\alpha_1$ becomes
maximal for intermediate values, $\theta_{\rm eff} \simeq 0.9$ and
$\theta_{\rm eff} \simeq 2.2$. In between these two values, $h_1$ is
mostly CP--odd, i.e. $\left| O^H_{13} \right| \geq 1/\sqrt{2}$; in the
remaining region $h_1$ is dominated by its CP--even components. 

Note that $\alpha_1$ is actually quite small, $\sim 0.1$, for
``maximal'' CP violation in the squark sector. The reason is that the
$(a, \phi_1)$ element (\ref{e13a}) of the Higgs mass matrix is small
here. The leading contribution to this element is proportional to the
quantity $R_t$ of eq.(\ref{e14b}), which is small if $A_t \mu {\em
e}^{i\xi}$ is purely imaginary, since the contribution $\propto
|\mu|^2$ is suppressed for our choice $\tanb=10$. As a result, this
element of the Higgs mass matrix goes through zero for $\theta_{\rm
eff} \simeq 0.53 \pi$. The quantity $R_t'$ remains large in this
region of parameter space, leading to a sizable $(a,\phi_2)$ element
of the Higgs mass matrix. However, its contribution to $\alpha_1$ is
suppressed by the relatively large difference between the diagonal
$(a,a)$ and $(\phi_2,\phi_2)$ elements.

\begin{figure}[htb]
\vspace*{-20mm}
\epsfig{file=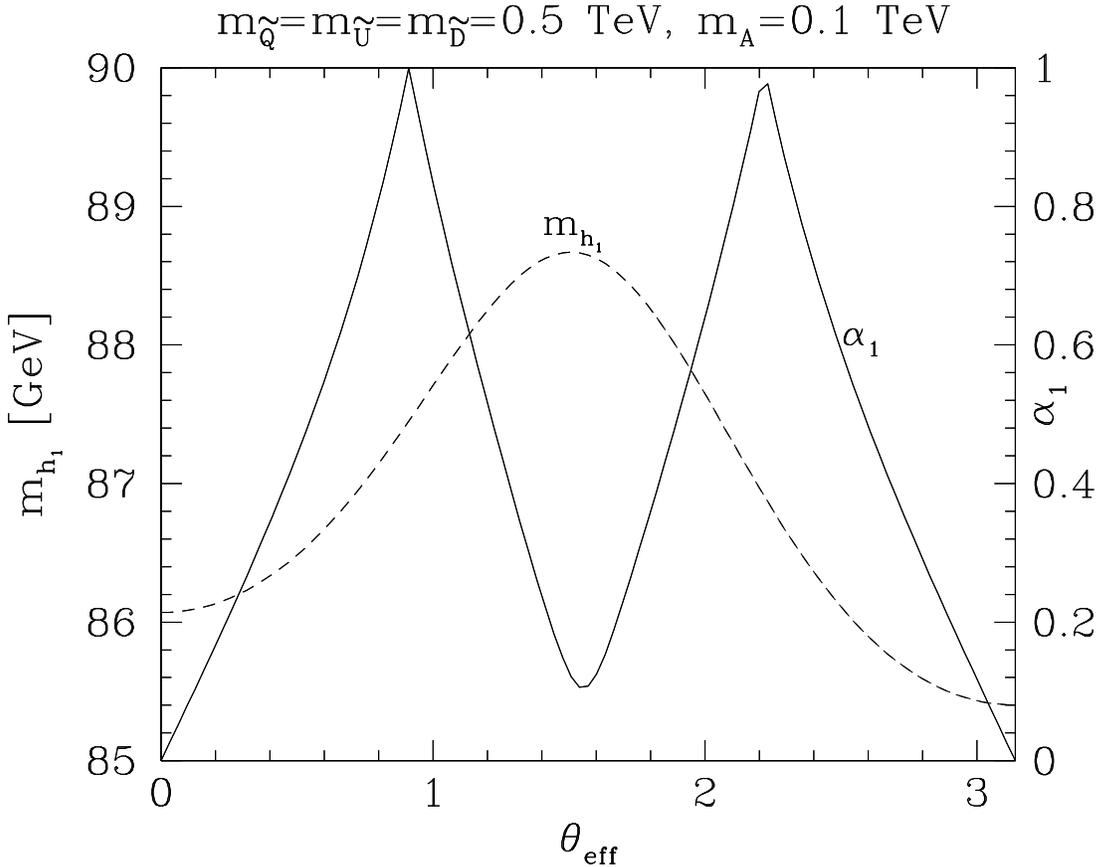,width=0.72\textwidth} 
\vspace*{-14mm}
\caption{\small The mass of the lightest neutral Higgs boson (dashed,
referring to the scale at the left) and the CP violating mixing angle
of this Higgs boson (solid, referring to the scale at the right) as a
function of the effective squark phase $\theta_{\rm eff} = \arg(A_t
\mu {\rm e}^{i\xi})$. We have fixed $m_{\wt Q} = m_{\wt U} = m_{\wt D}
= m_{\tilde g} = 500$ GeV and $m_A = 100$ GeV; the values of the other 
parameters are as in Fig.~1.}
\end{figure}

\section*{4) Summary and conclusions}

In this paper we calculated quark and squark loop corrections to the
mass matrix of the neutral Higgs bosons of the MSSM using the effective
potential method. We allowed for CP violating phases in the squark
sector as well as arbitrary splitting between squark masses. We also
for the first time included mixed weak gauge--Yukawa contributions, as
well as leading 2--loop SUSY--QCD contributions, in the calculation of
induced CP violation in the Higgs sector; the latter have been
absorbed into properly defined running quark masses, following
ref.\cite{2}. Our formulas can be used to accurately calculate the
masses and mixing angles of the MSSM Higgs bosons for all values of
the (possibly complex) parameters describing the third generation
squark mass matrices.

When comparing our results with earlier expressions \cite{5} that are
valid for relatively small squark mass splitting we found surprisingly
good agreement for the Higgs spectra even in cases with very large
squark mass splitting. However, we also found $\sim 30\%$
discrepancies in the predictions of CP violating Higgs mixing angles
if the difference between squark masses is large. Similarly, we found
that the mixed weak gauge--Yukawa and 2--loop SUSY--QCD contributions
do not change the predictions for the Higgs spectrum very much, but
can change the prediction for CP violating Higgs mixing angles by
$\sim 20\%$.

Finally, we showed that maximal mixing between CP--even and CP--odd
Higgs states remains possible for reasonable choices of the free
parameters. Note that this usually does not happen when CP violation
in the squark sector is maximal, i.e. CP violating Higgs mixing angles
reach their maximum for intermediate values of the effective CP
violating phase in the scalar top mass matrix (away from
$\pi/2$). Since the loop--induced CP violation in the MSSM Higgs
sector can be very large, an accurate treatment of this effect, as
described in this paper, is required for the interpretation of
searches for Higgs bosons at LEP and elsewhere.

\bigskip

\noindent {\bf Acknowledgements:} \\
S.Y.C. wishes to acknowledge financial support of the 1997 Sughak
program of the Korea Research Foundation. The work of M.D. was
supported in part by the ``Sonderforschungsbereich 375--95 f\"ur
Astro--Teilchenphysik'' der Deutschen Forschungsgemeinschaft.

\end{document}